\documentclass[preprint,showpacs,preprintnumbers,amsmath,amssymb]{revtex4}

\usepackage{graphicx}
\usepackage{dcolumn}
\usepackage{bm}

\begin{document}

\preprint{}

\title{Topological structure of the many vortices solution in Jackiw-Pi model}
\author{Xi-Guo Lee $^{1,2}$}
\author{Zi-Yu Liu  $^{1,3}$}
\email{liuziyu@impcas.ac.cn}
\author{Yong-Qing Li $^{1,3}$}
\author{Peng-Ming Zhang $^{1,2}$}
\affiliation{$^{1}$Institute of Modern Physics, Chinese Academy of
Science, P.O. Box 31, Lanzhou 730000, People's republic of china}
\affiliation{$^{2}$Center of Theoretical Nuclear Physics, National
Laboratory of Heavy Ion Collisions, Lanzhou 730000, China}
\affiliation{$^{3}$Graduate School, Chinese Academy of Science,
Beijing 100049, People's republic of china}


\begin{abstract}
By using the gauge potential decomposition, we discuss the
self-dual equation and its solution in Jackiw-Pi model. We obtain
a new concrete self-dual equation and find relationship between
Chern-Simons vortices solution and topological number which is
determined by Hopf indices and Brouwer degrees of $\Psi$-mapping.
To show the meaning of topological number we give several figures
with different topological numbers. In order to investigate the
topological properties of many vortices, we use $5$ parameters(two
positions, one scale, one phase per vortex and one charge of each
vortex) to describe each vortex in many vortices solutions in
Jackiw-Pi model. For many vortices, we give three figures with
different topological numbers to show the effect of the charge on
the many vortices solutions. We also study the quantization of
flux of those vortices related to the topological numbers in this
case.

\end{abstract}
\maketitle

\section{introduction}
Chern-Simons theories based on secondary characteristic classes
discovered in Ref.\cite{cs01} exhibit many interesting and
important physical properties. In the early 1980, the first
physical applications of the Chern-Simons form called
topologically massive gauge theory was advanced by J.
Schonfeld\cite{ho01}, and many topological invariants of knots and
links discovered in the 1980s could be reinterpreted as
correlation functions of Wilson loop operators in Chern-Simons
theory\cite{wt03}. Moreover, for gauge theories and gravity in
three-dimensions, they can appeared as natural mass terms and will
lead to a quantized coupling constant as well as a mass after
quantization\cite{dj02}. They have also found applications to a
lot of physical problems, such as particle physics, quantum Hall
effect, quantum gravity and string
theory\cite{ykf,ba98,jp01,re01,qh01,re02,re03,re04,re05,wt01,wt02}.
 Chern-Simons term acquire dynamics via
coupling to other fields\cite{ah01,ba98}, and get multifarious
gauge theory, non-relativistic Chern-Simons theory supports
vortices solutions, these static solutions can
 be obtained when their Hamiltonian was minimal. Vortices and their dynamics are interesting objects
 to be studied\cite{ah01,pd02,pd23,pd33,pd43,duth}. R. Jackiw and S-Y. Pi
considered a gauged, nonliner Sch\"{o}dinger equation in two
spatial dimensions, with describes non-relativistic  matter
interacting with Chern-Simons gauge fields. Then they find
explicit static, self-dual solutions which satisfies the Liouville
equation and got an $n$-solitons solutions depends on $4n$
parameters(two positions, one scale, one phase per
solition)\cite{ba98,jp01}. P. A. Horv$\acute{\mathrm{a}}$thy
proved that the solution depends on $4N$ parameters without the
use of an index theorem and the flux quantization\cite{pa01}, and
indicated the regular solutions with finite degree only arise for
rational functions, the topological degree of those solutions is
the common number of their zeros and poles on the Riemann
sphere\cite{pa02}.

In this paper, by using the gauge potential
decomposition\cite{dg01,du01,sa01,li04,li05}, we will discuss
topological structure of the self-dual solution in Jackiw-Pi
model. We will look for complete many vortices solution from the
self-dual equation and set up the relationship between the many
vortices solution and topological number which is determined by
Hopf indices and Brouwer degrees. We will give several figures to
show the effects of the topological number on vortices. We also
study the quantization of the flux of those vortices.

\section{the topological number of Self-dual vortex in Jackiw-Pi Model}
In this section, based on the self-dual equation, we will look for
complete vortex solution in Jackiw-Pi model\cite{ba98} making use
of the decomposition of gauge potential. The Abelian Jackiw-Pi
model in nonlinear Schr\"odinger systems is\cite{ba98,jp01}
\begin{eqnarray}
\pounds_{jp}=\frac{k}{2}\epsilon^{\mu\nu\lambda}A_{\mu}\partial_{\nu}A_{\lambda}+i\hbar\Psi^{*}D_{0}\Psi-\frac{\hbar^{2}}{2m}|\mathbf{D}\Psi|^{2}+\frac{g}{2}(\Psi^{*}\Psi)^{2}.
\end{eqnarray}
[Here relativistic notation with the metric diag is (1,-1,-1) and
$x^{\mu}=(ct,\mathbf{r})$.] Where
$\mathbf{D}=\mathbf{\nabla}-i\frac{e}{\hbar c}\mathbf{A}$ and
$\Psi$ is "matter" field, the first term is the Chern-Simons
density, which is not gauge invariant. Also $m$ is the mass
parameter, $A_{\mu}$ is gauge potentials, $g$ governs the strength
of nonlinearity, $\kappa$ controls the Chern-Simons addition and
provides a cutoff at large distance greater than
$\frac{1}{\kappa}$ for gauge-invariant electric and magnetic
fields, which can be written as $\mathbf{E}=-\nabla
\mathbf{A}^{0}-(\frac{1}{c})\partial_{t}\mathbf{A}$, and
$\mathbf{B}=\mathbf{\nabla}\times\mathbf{A}$. Thus the
Chern-Simons terms gives rise to massive, yet gauge-invariant
"electrodynamics". The last term represents a self-coupling
contact term of the type commonly found in nonlinear Schr¡§odinger
systems. The magnetic fields $B$ satisfies
\begin{eqnarray}
B=\epsilon^{ij}\partial_{i}A^{j}=-\frac{e}{k}\rho,
\end{eqnarray}
where
\begin{eqnarray}
\rho=\Psi^{*}\Psi,
\end{eqnarray}
 with $g=\mp\frac{e^{2}\hbar}{mck}$, and sufficiently well-behaved
fields so that the integral over all space of
$\nabla\times\mathbf{J}$ vanishes, the energy is
\begin{eqnarray}
H=\int d\mathbf{r}\mathcal{H}=\frac{\hbar^{2}}{2m}\int
d\mathbf{r}|(D_{1}\pm iD_{2})\Psi|^{2}.
\end{eqnarray}
This is non-negative and vanishes. So it is obvious that $\Psi$
satisfies a self-dual equation
\begin{eqnarray}
D_{1}\Psi=\mp iD_{2}\Psi.
\end{eqnarray}
To solve Eq.(5),we note that when $\Psi$ is decomposed into two
scalar fields
\begin{eqnarray}
\Psi=\Psi_{1}+i\Psi_{2}.
\end{eqnarray}

We can define a unit vector field $\mathbf{n}$ as follows
\begin{eqnarray}
n^{a}=\frac{\Psi_{a}}{(\Psi^{*}\Psi)^{\frac{1}{2}}},a=1,2.
\end{eqnarray}
It is easy to prove that $\mathbf{n}$ satisfies the constraint
conditions
\begin{eqnarray}
n^{a}n^{a}=1,    n^{a}dn^{a}=0.
\end{eqnarray}
From Eq.(5), and making use of the decomposition of U(1) gauge
potential in terms of the two-dimensional unit vector field
$\mathbf{n}$\cite{li05}, we can obtain
\begin{eqnarray}
A^{i}=\frac{\hbar
c}{e}(\epsilon^{ab}n^{a}\partial_{i}n^{b}\pm\frac{1}{2}\epsilon^{ij}\partial_{j}\ln\rho).
\end{eqnarray}
From Eq.(2) and Eq.(9) we get
\begin{eqnarray}
B=\frac{\hbar
c}{e}\epsilon^{ij}\epsilon^{ab}\partial_{i}n^{a}\partial_{j}n^{b}\pm\frac{\hbar
c}{2e}\nabla^{2}\ln\rho,
\end{eqnarray}
i.e.
\begin{eqnarray}
-\frac{e}{k}\rho=\frac{\hbar
c}{e}\epsilon^{ij}\epsilon^{ab}\partial_{i}n^{a}\partial_{j}n^{b}\pm\frac{\hbar
c}{2e}\nabla^{2}\ln\rho.
\end{eqnarray}
This equation can be rewritten as
\begin{eqnarray}
\nabla^{2}\ln\rho=\pm\frac{2e^{2}}{\hbar c
k}\rho\pm2\epsilon^{ij}\epsilon^{ab}\partial_{i}n^{a}\partial_{j}n^{b},
\end{eqnarray}
with the help of the $\phi$ -mapping method\cite{du01,sa01},
Eq.(12) can be written as
\begin{eqnarray}
\nabla^{2}\ln\rho=\pm\frac{2e^{2}}{\hbar c
k}\rho\pm4\pi\delta^{2}(\mathbf{\Psi})J\left(\frac{\mathbf{\Psi}}{\mathbf{r}}\right),
\end{eqnarray}
in which the vector field $\mathbf{\Psi}$ is defined by
$\mathbf{\Psi}=(\Psi_{1},\Psi_{2})$, and
$J\left(\frac{\mathbf{\Psi}}{\mathbf{r}}\right)$ is Jacobian
\begin{eqnarray}
J\left(\frac{\mathbf{\Psi}}{\mathbf{r}}\right)=\frac{1}{2}\epsilon^{ab}\epsilon^{ij}\frac{\partial\Psi_{a}}{\partial
x^{i}}\frac{\partial\Psi_{b}}{\partial x^{j}},       i,j=1,2.
\end{eqnarray}
When $\rho\neq0$, Eq.(13) will be the Liouville equation,
\begin{eqnarray}
\nabla^{2}\ln\rho=\pm\frac{2e^{2}}{\hbar c k}\rho,
\end{eqnarray}
as we all know, the Eq.(15) has the general real solution as
follows
\begin{equation}
\rho(\mathbf{r})=\frac{4\hbar
c\kappa}{e^{2}}\frac{|f^{'}(z)|^{2}}{[1+|f(z)|^{2}]^{2}},
\end{equation}
in which
\begin{eqnarray}
\mathbf{r}=(r\cos\theta,r\sin\theta),
\end{eqnarray}
where$z=re^{i\theta}$ and $f(z)$ is an arbitrary
function\cite{ba98}.

Because $\rho$ is the charge density of the vortex, it must be
positive, the Liouville equation is
\begin{eqnarray}
\nabla^{2}\ln\rho=-\frac{2e^{2}}{\hbar c|\kappa|}\rho,
\end{eqnarray}
so the Eq.(16) should be
\begin{eqnarray}
\rho(\mathbf{r})=\frac{4\hbar
c|\kappa|}{e^{2}}\frac{|f^{'}(z)|^{2}}{[1+|f(z)|^{2}]^{2}},
\end{eqnarray}
and the Eq.(13) can rewritten as
\begin{eqnarray}
\nabla^{2}\ln\rho=-\frac{2e^{2}}{\hbar c
|\kappa|}\rho-(sgn\kappa)4\pi\delta^{2}(\mathbf{\Psi})J\left(\frac{\mathbf{\Psi}}{\mathbf{r}}\right).
\end{eqnarray}
It is a Liouville equation with a $\delta-$function on its right
side. For $\mathbf{r}=0$,this equation is also right. To show the
meaning of the $\delta-$function of this equation, we will
integrate Eq.(20) in section $III$, and discuss its singular
point. For one vortex
\begin{eqnarray}
f(z)=\left(\frac{c}{z-z_{0}}\right)^N,
\end{eqnarray}
in which $c=r_{c}e^{i\theta_{c}}$ and $z_{0}=r_{0}e^{i\theta_{0}}$
is a complex constant, so there are $5$ real parameters involved
in this solution : $2$ real parameters $z_{0}$ describing the
locations of the vortices, $2$ real parameters $c$ corresponding
to the scale and phase of each vortex, $1$ real parameters $N$
describing the charge of the vortex, and it is easy to obtain the
radially symmetric solutions\cite{we01}
\begin{eqnarray}
\rho=\frac{\hbar c}{e^{2}}{4|\kappa|N^2\over
r_{c}^2}{\left(\frac{r-r_{0}}{r_{c}}\right)^{2N-2}\over
\left(1+\left(\frac{r-r_{0}}{r_{c}}\right)^{2N}\right)^2},
\end{eqnarray}
where
\begin{eqnarray}
N=-\frac{\kappa}{|\kappa|}Q+1 \\
Q=\beta\eta,
\end{eqnarray}
where $Q$ is topological number of the vortex, those number is
determined by Hopf indices and Brower degree of $\Psi$.
Particularly, $\rho$ is invariant when change $N$ to $-N$, see
Figure [1] and Figure [2] for a plot of the one vortex case, in
this case, the center of the vortex is $z_{0}=3+3i$. In order to
study the shape of the vortex, we slice off Figure[2] through it's
center, and define the hight and radius of the vortex, as is shown
in Figure[3], we can obtain
\begin{eqnarray}
r_{v}=r_{c}\left(1-\frac{2}{-\frac{\kappa}{|\kappa|}Q+1}\right)^{\frac{1}{-2\frac{\kappa}{|\kappa|}Q+2}}.
\end{eqnarray}
Figure [5] shows the values of the radius of the vortex as $Q$ is
varied.

The height of the vortex
\begin{eqnarray}
\rho_{max}=\frac{\hbar
c|\kappa|}{r_{c}^{2}e^{2}}\left(\frac{\frac{\kappa}{|\kappa|}Q}{\frac{\kappa}{|\kappa|}Q-2}\right)^{\frac{1}{\frac{\kappa}{|\kappa|}Q-1}}\left(\left(\frac{\kappa}{|\kappa|}Q-1\right)^{2}-1\right).
\end{eqnarray}
Figure [4] shows the values of the height of the vortex as $Q$ is
varied.

\section{the topological structure of the many vortices solution and their magnetic flux }
In this section, making use of Eq.(20), we will discuss the
topological structure of the many vortices solution, then we will
study the magnetic flux of the vortices. The meromorphic function
$f(z)$ yields a regular many vortices solution with finite
magnetic flux if and only if $f(z)$ is a rational function,
\begin{eqnarray}
f(z)=\frac{P(z)}{T(z)},
\end{eqnarray}
subject to
\begin{eqnarray}
degP < degT.
\end{eqnarray}
In particular, when all roots of $T(z)$ are simple, $f(z)$ can be
developed in to partial fractions\cite{pa01},
\begin{eqnarray}
f(z)=\sum_{a=1}^{M}\frac{c_{a}}{z-z_{a}},
\end{eqnarray}
in which
\begin{eqnarray}
z_{a}=r_{a}e^{i\theta_{a}},c_{a}=r_{0a}e^{i\theta_{0a}},a=1,2,...M.
\end{eqnarray}
So there are $4M$ real parameters involved in this solution : $2M$
real parameters $r_{a}$ and $\theta_{a}$ describing the locations
of the vortices, $2M$ real parameters $r_{0a}$ and $\theta_{0a}$
corresponding to the scale and phase of each vortex. However,
there is non evidence to believe the charge of each vortex equals
to $1$, in order to study the charge and topological structure of
each vortex in this solution, we suppose the charge of vortex
$z_{a}$ is $N_{a}$(vortex $z_{a}$ is the vortex whose center is
$z_{a}$), the we can rewrite $f(z)$ as
\begin{eqnarray}
f(z)=\sum_{a=1}^{M}\left(\frac{c_{a}}{z-z_{a}}\right)^{N_{a}},
\end{eqnarray}
this describing $M$ separated $charge-N_{a}$ vortices, and
$N_{a}>0$ because of $degP < degT$, then we add $M$ real
parameters $N_{a}$ in our solution, these parameters describing
the charge of each vortex and in the following text we will see
that $N_{a}$ is relates to the topological number of each vortex.

Under the radially symmetric, $\nabla^{2}\ln\rho$ can be expressed
as
\begin{eqnarray}
\nabla^{2}\ln\rho=\frac{\partial^{2}}{\partial
^{2}r}\ln\rho+\frac{1}{r}\partial_{r}\ln\rho.
\end{eqnarray}
Integrating Eq.(20)
\begin{eqnarray}
\int\nabla^{2}\ln\rho d\mathbf{r}=\int\left[-\frac{2e^{2}}{\hbar c
|\kappa|}\rho-(sgn\kappa)4\pi\delta^{2}(\mathbf{\Psi})J\left(\frac{\mathbf{\Psi}}{\mathbf{r}}\right)\right]d\mathbf{r}.
\end{eqnarray}
The Eq.(33) can be rewritten as
\begin{eqnarray}
\int_{\mathbf{r}_{a}}^{\mathbf{r}_{a}+\mathbf{r}_{\epsilon}}\nabla^{2}\ln\rho
d\mathbf{r}=-(sgn\kappa)4\pi\int_{\mathbf{r}_{a}}^{\mathbf{r}_{a}+\mathbf{r}_{\epsilon}}\delta^{2}(\mathbf{\Psi})J\left(\frac{\mathbf{\Psi}}{\mathbf{r}}\right)d\mathbf{r},
\end{eqnarray}
where
$\mathbf{r}_{\epsilon}=(r_{\epsilon}\cos\theta,r_{\epsilon}\sin\theta),0\leq\theta\leq
2\pi$, and $r_{\epsilon}$ is a infinitesimal scalar, so
$\int_{\mathbf{r}_{a}}^{\mathbf{r}_{a}+\mathbf{r}_{\epsilon}}$ is
an integral in the infinitesimal region neighbouring the
$\mathbf{r}_{a}$ point. The left side of this equation is
\begin{eqnarray}
\int_{\mathbf{r}_{a}}^{\mathbf{r}_{a}+\mathbf{r}_{\epsilon}}\nabla^{2}\ln\rho
d\mathbf{r}=4\pi(N_{a}-1).
\end{eqnarray}
Suppose that the vector field $\Psi^{a}$ possess $M$ isolated
zeros which is in $\mathbf{r}=\mathbf{r}_{a}$, according to the
$\delta$-function theory\cite{delt}, $\delta^{2}(\mathbf{\Psi})$
can be expressed by
\begin{equation}
\delta^{2}(\mathbf{\Psi})=\sum_{a=1}^{M}\frac{\beta_{a} }{\mid
J\left(\frac{\mathbf{\Psi}}{\mathbf{r}}\right)\mid_{\mathbf{r}=\mathbf{r}_{a}}}\delta^{2}(\mathbf{r}-\mathbf{r}_{a}),
\end{equation}
and then we can obtain
\begin{eqnarray}
\int_{\mathbf{r}_{a}}^{\mathbf{r}_{a}+\mathbf{r}_{\epsilon}}4\pi\delta^{2}(\mathbf{\Psi})J\left(\frac{\mathbf{\Psi}}{\mathbf{r}}
\right)d\mathbf{r}=4\pi
\int_{\mathbf{r}_{a}}^{\mathbf{r}_{a}+\mathbf{r}_{\epsilon}}
\beta_{a}\eta_{a}\delta^{2}(\mathbf{r}-\mathbf{r}_{a})d\mathbf{r},
\end{eqnarray}
where $\beta_{a}$ is positive integer (the Hopf index of the zero
point) and $\eta_{a}$, the Brouwer degree of the vector field
$\mathbf{\Psi}$,
\begin{eqnarray}
\eta_{a}=sgnJ\left(\frac{\mathbf{\Psi}}{\mathbf{r}}\right)|_{\mathbf{r}=\mathbf{r}_{a}}=\pm1.
\end{eqnarray}
The meaning of the Hopf index $\beta$ is that while $\mathbf{r}$
covers the region neighbouring the $z_{a}$ point once, the vector
field $\mathbf{\Psi}$ covers the corresponding region $\beta_{a}$
times. Hence, $\beta_{a}$ and $\eta_{a}$ are the topological
number which shows the topological properties of the vortex
solution. We have
\begin{eqnarray}
\delta^{2}(\mathbf{\Psi})J\left(\frac{\mathbf{\Psi}}{\mathbf{r}}\right)=\beta_{a}\eta_{a}\delta^{2}(\mathbf{r}-\mathbf{r}_{a}).
\end{eqnarray}
If we define the topological number $Q_{a}$ of the vortex whose center is $z_{a}$
as
\begin{eqnarray}
Q_{a}=\int_{\mathbf{r}_{a}}^{\mathbf{r}_{a}+\mathbf{r}_{\epsilon}}\delta^{2}(\mathbf{\Psi})J\left(\frac{\mathbf{\Psi}}{\mathbf{r}}\right)d\mathbf{r}=\beta_{a}\eta_{a},
\end{eqnarray}
from Eq.(34) we can get
\begin{eqnarray}
N_{a}-1=-(sgn\kappa)Q_{a},
\end{eqnarray}
in which $N_{a}-1\geq0$, so $\kappa$ must satisfied
$sgn\kappa=-sgn J$ and we can set $-(sgn\kappa)Q_{a}=|Q_{a}|$,
then the total charge of those vortices
\begin{eqnarray}
N=degT=\sum_{a=1}^{M}N_{a}=\sum_{a=1}^{M}|Q_{a}|+M=-(sgn\kappa)Q+M,
\end{eqnarray}
in which $Q$ is the total topological number which can be defined
as
\begin{eqnarray}
Q=\sum_{a=1}^{M}Q_{a}.
\end{eqnarray}
 Substituting Eq.(41) into Eq.(31), we can obtain
\begin{eqnarray}
f(z)=\sum_{a=1}^{M}\left(\frac{c_{a}}{z-z_{a}}\right)^{|Q_{a}|+1},
\end{eqnarray}
it is obviously that Eq.(44)  and Eq.(19)is the solution of
Eq.(20). On the other hand, this means vortices density $\rho$
relates to its topological number $Q_{a}=\beta_{a}\eta_{a}$. We
now see that $N_{a}$ must be an integer.

We can see the point $r_{a}$ is a singular point of the field
$\mathbf{\Psi}$, and it is non-degenerate. In our case, we can
list the possible four types of non-degenerate singular points in
two dimensions\cite{book}:

$1)$a center with index $1$(when both eigenvalues are purely
imaginary);

$2)$a node with index $1$(eigenvalues real and the same sign);

$3)$a focal point index $1$(complex conjugate eigenvalues, not
purely imaginary);

$4)$a saddle point of index $-1$(eigenvalues real and the opposite
sign).

Information about singular piont of a vector field is of grate
importance for obtaining a qualitative picture of the behavior of
the integral trajectories of the field.

If we note the unit magnetic flux $\Phi_{0}=\frac{2\pi\hbar
c}{e}$, one can get
\begin{eqnarray}
\int_{0}^{\infty}
\mathbf{B}d\mathbf{r}=-(sgn\kappa)\frac{2Nhc}{e}=-2(sgn\kappa)\Phi_{0}\left[-(sgn\kappa)Q+M\right],
\end{eqnarray}
from this equation we know the magnetic flux is quantized. When
the total topological number equal to zero, the magnetic flux of
this vortex is
\begin{eqnarray}
\Phi=\int_{0}^{\infty}
\mathbf{B}d\mathbf{r}=-2(sgn\kappa)M\Phi_{0}.
\end{eqnarray}

For example, when set $M=1$ in Eq.(31), we can get the one vortex
solution Eqs.(22). When set $M=2$ in Eq.(31), we can get the two
vortices solution with $z_{1}=-3,z_{2}=3,c_{1}=c_{2}=1$, i.e.
\begin{eqnarray}
f(z)=\left(\frac{1}{z+3}\right)^{|Q_{1}|+1}+\left(\frac{1}{z-3}\right)^{|Q_{2}|+1},
\end{eqnarray}
with Eq.(19), we can give the solution of two vortices. See Figure
[4] for a plot of the two vortex case.

When set $M=3$ in Eq.(31), we can get the three vortices solution
with $z_{1}=-3-\sqrt{3}i,z_{2}=3-\sqrt{3}i, z_{3}=2\sqrt{3}i,
c_{1}=c_{2}=c_{3}=1$, it is as
\begin{eqnarray}
f(z)=\left(\frac{1}{z-z_{1}}\right)^{|Q_{1}|+1}+\left(\frac{1}{z-z_{2}}\right)^{|Q_{2}|+1}+\left(\frac{1}{z-z_{3}}\right)^{|Q_{3}|+1},
\end{eqnarray}
with Eq.(19), we can obtain the solution of three vortices. See
Figure [7] for a plot of the three vortices case and Figure[8] for
the four vortices case. Note the ring-like form of the magnetic
field for these Chern-Simons vortices, as the magnetic field is
proportional to $\rho$, $\mathbf{B}$ vanishes where the field
$\Psi$ vanishes. From those figures we can see the shape of those
vortices is different when their topological numbers is different.
So it is not enough to use $4M$ parameters to describe many
vortices, we must induct a charge parameter $N$, from Figure[4]
and Figure[5] we can see, the hight and radius of the vortex
depends on $Q$.

\section{Conclusions}
In this paper, We discuss the self-dual equation and its solution
in Jackiw-Pi model By using the gauge potential decomposition and
$\Psi$-mapping method, we get a Liouville equation with a $\delta$
function, then we also obtain the solution of this equation, and
the $\delta$ function will not change the character of the
solution when $\rho\neq0$. We add $M$ parameters to the
$M$-vortices solution, those parameters describing the charge of
each vortex, i.e. use $5M$ parameters(two positions, one scale,
one phase per vortex and one charge per vortex) to describe
$M$-vortices solutions in Jackiw-Pi model. We studied the
topological structure of Chern-Simons vortices in Jakiw-Pi model
by calculate the integral of the Liouville equation, and find the
charges of those vortices are determined by the topological
numbers of those vortices, those topological numbers are
determined by Hopf indices and Brouwer degrees of $\Psi$-mapping,
the total topological number of these vortices is conservational.
We also give some figures with different topological numbers to
show the relationship between the shape of those vortex and
topological numbers. In many vortices solution, in order to show
the shape of those vortices is different when only their
topological number is different, we show some vortices only with
different positions and different topological numbers in many
vortices figures(see Figure[6];[7];[8]). We also find the
relationship between the quantization of the flux and the
topological numbers from the integral value of the solution in the
whole space. However, the flux is non-vanish when the topological
number equals to zero. So does the the angular momentum.

\section{acknowledgments}
We thank professor P. A. Horv$\acute{\mathrm{a}}$thy for his
important information. This work was supported by the CAS
Knowledge Innovation Project (KJX2-SW-N16;KJCX3-SYW-N2) and
Science Foundation of China (10435080, 10275123).

\section{references}

 \newpage

\begin{figure}[ht]
  \begin{center}
    \rotatebox{0}{\includegraphics*[width=0.7\textwidth]{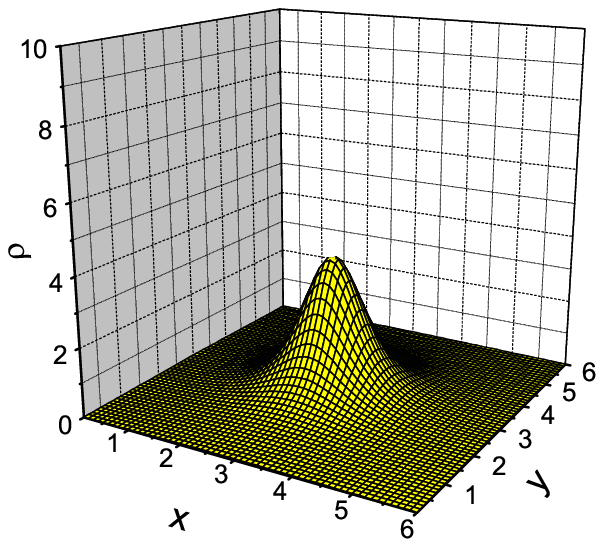}}
    \caption{Density $\rho$ for solution (22) representing one vortex with $Q=0$.}
  \end{center}
\end{figure}

\begin{figure}[ht]
  \begin{center}
    \rotatebox{0}{\includegraphics*[width=0.7\textwidth]{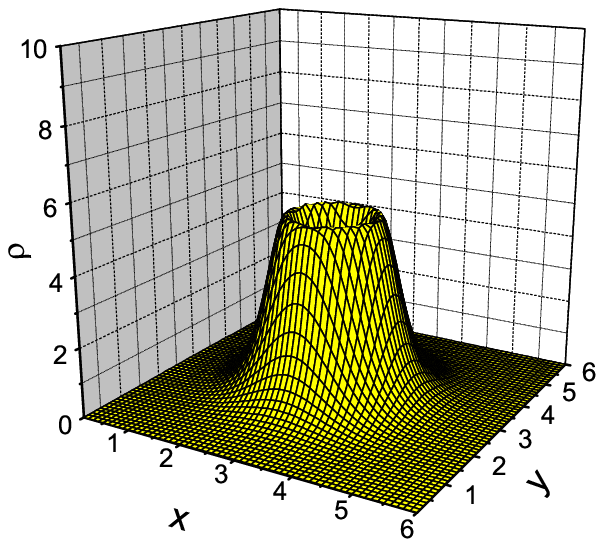}}
    \caption{Density $\rho$ for solution (22) representing one vortex with $Q=1$.}
  \end{center}
\end{figure}

\begin{figure}[ht]
  \begin{center}
    \rotatebox{0}{\includegraphics*[width=0.7\textwidth]{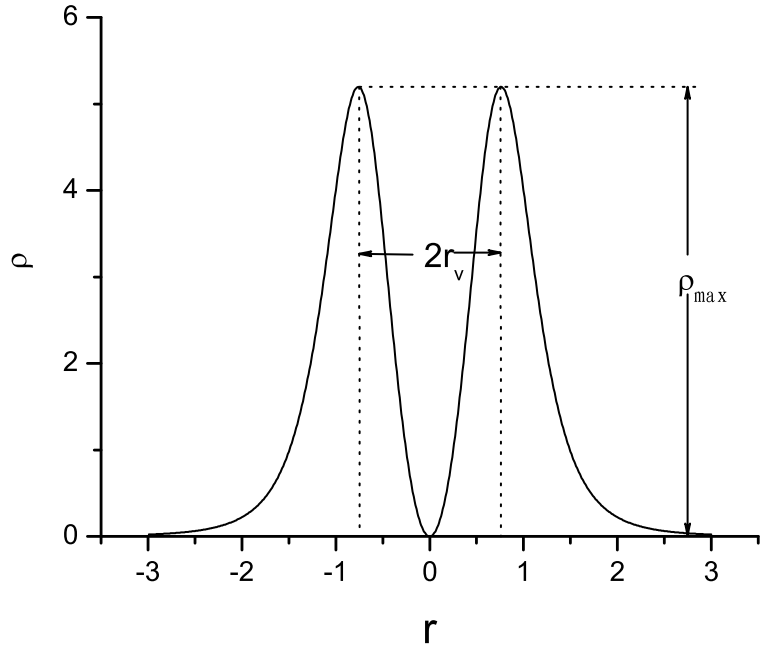}}
    \caption{The section plane of Figure[2], this section plan includes the center of this vortex, as is shown in this figure, we can define the hight $\rho_{max}$ and radius $r_{v}$ of the vortex.}
  \end{center}
\end{figure}

\begin{figure}[ht]
  \begin{center}
    \rotatebox{0}{\includegraphics*[width=0.7\textwidth]{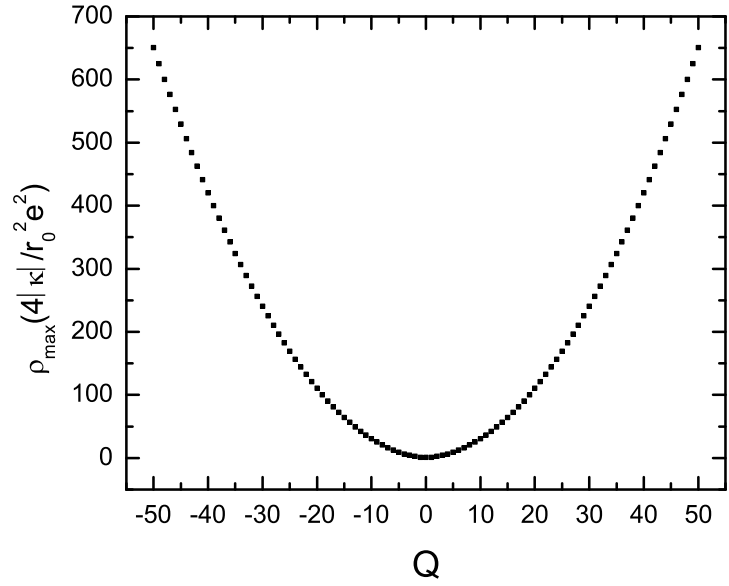}}
    \caption{The values of the height of the vortex as $Q$ is varied$(\hbar=c=1)$.}
  \end{center}

\end{figure}
\begin{figure}[ht]
  \begin{center}
    \rotatebox{0}{\includegraphics*[width=0.7\textwidth]{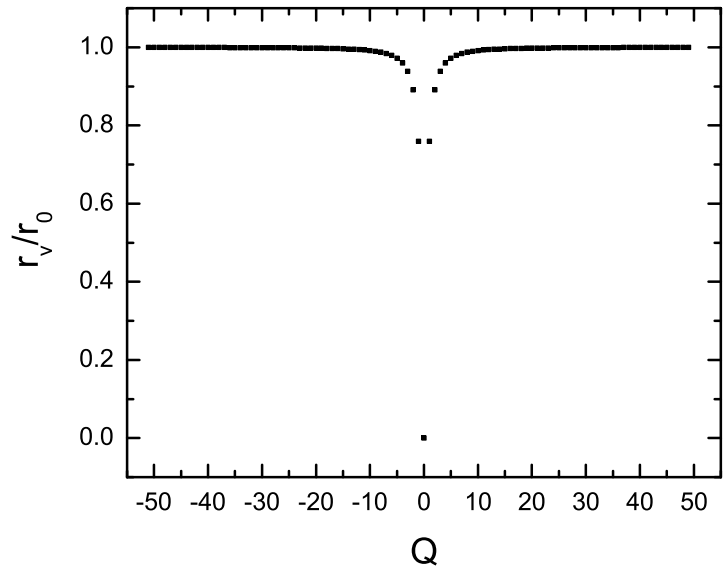}}
    \caption{The values of radius of the vortex as $Q$ is varied$(\hbar=c=1)$.}
  \end{center}
\end{figure}

\begin{figure}[ht]
  \begin{center}
    \rotatebox{0}{\includegraphics*[width=0.7\textwidth]{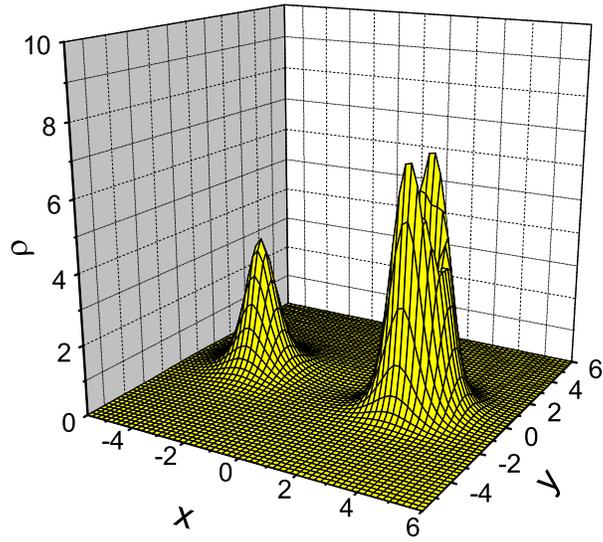}}
    \caption{Density $\rho$ for solution (47) representing two separated vortices with $Q_{1}=0,Q_{2}=1$.}
  \end{center}
\end{figure}

\begin{figure}[ht]
  \begin{center}
    \rotatebox{0}{\includegraphics*[width=0.7\textwidth]{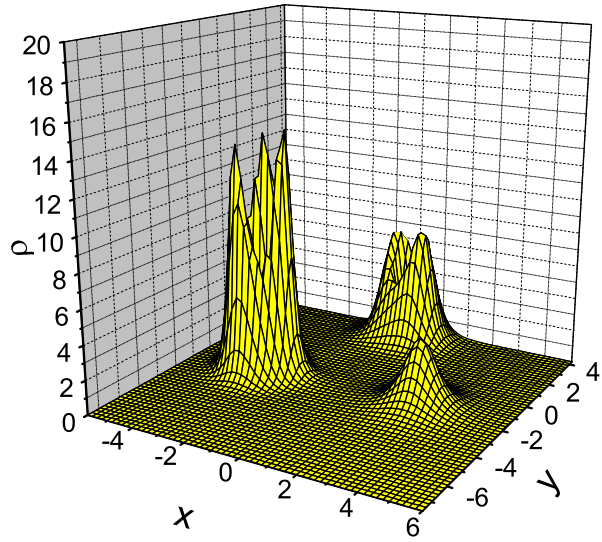}}
    \caption{Density $\rho$ for solution (48) representing three separated vortices with $Q_{1}=0,Q_{2}=1,Q_{3}=2$.}
  \end{center}
\end{figure}

\begin{figure}[ht]
  \begin{center}
    \rotatebox{0}{\includegraphics*[width=0.7\textwidth]{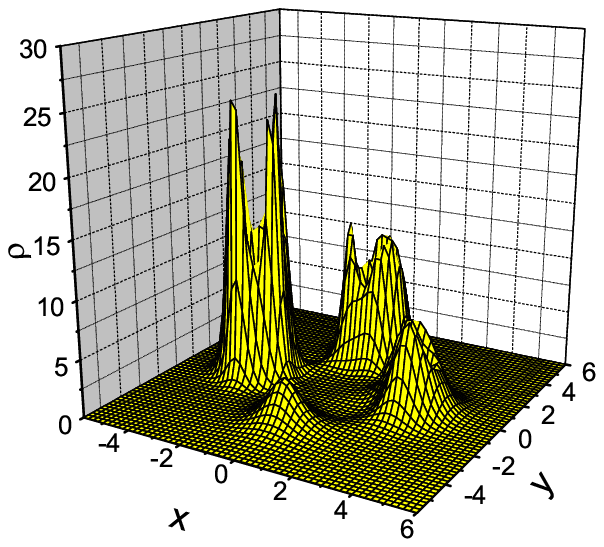}}
    \caption{Density $\rho$ representing four separated vortices with $Q_{1}=0,Q_{2}=1,Q_{3}=2,Q_{4}=3$.}
  \end{center}
\end{figure}

\begin{thebibliography}{99}
\baselineskip 0.05in
\bibitem{cs01}S. S. Chern and J. Simons, Proc. Nat. Acad. Sci. USA 68(4), 791 (1971);
              S. S. Chern and J. Simon,Ann. Math. 99, 48 (1974).
\bibitem{ho01}J. Schonfeld, Nucl. Phys. B185, 157 (1981).
\bibitem{wt03}E. Witten, Commun. Math. Phys. 121, 351 (1989).
\bibitem{dj02}S. Deser, R. Jackiw and S. Templeton, Phys. Rev. Lett. 48, 975 (1982).
              S. Deser, R. Jackiw and S. Templeton, Ann. of Physics 140, 372 (1982).
\bibitem{ykf} R. Jackiw and E. J. Weinberg, Phys. Rev. Lett. 64, 2234 (1990).
\bibitem{ba98}R. Jackiw and S. Y. Pi, Phys. Rev. Lett. 64, 2969 (1990).
\bibitem{jp01}R. Jackiw and S. Y. Pi,  Phys. Rev.D 42, 3500 (1990).
\bibitem{re01}J. Hong, Y. Kim and P. Y. Pac, Phys. Rev. Lett. 64, 2230 (1990).
\bibitem{qh01}S. M. Girvin and A.H. MacDonald, Phys. Rev. Lett. 58, 1252
              (1987).
\bibitem{re02}S.C. Zhang, T.H. Hansson and S. Kivelson, Phys. Rev. Lett. 62, 82 (1988).
\bibitem{re03}V. Kalmeyer and R. B. Laughlin, Phys. Rev. Lett. 59, 2095 (1987);
\bibitem{re04}R. B. Laughlin, Phys. Rev. Lett. 60, 2677 (1988).
\bibitem{re05}A. Achucarro and P.K. Townsend , Phys. Lett. B 180, 89 (1986) .
\bibitem{wt01}E. Witten, Chern¨CSimons gauge theory as a string theory, in:
              The Floer Memorial Volume, in: Progress in Mathematics, vol. 133,
              Birkhauser, Boston, MA, 1995, pp. 637¨C678. arXiv:hep-th/9207094
\bibitem{wt02}E. Witten,  Nucl. Phys. B 311, 46 (1988).
\bibitem{ah01}A. A. Abrikosov, Sov.Phys. JETP 5,1174(1957).
\bibitem{pd02}Bogomol¡¯nyi, E.B.: The stability of classical solutions. Sov. J. Nucl. Phys. 24, 449 (1976).
\bibitem{pd23}H. Nielsen and P. Olesen, Nucl. Phys. B 61, 45 (1973).
\bibitem{pd33}H. J. de Vega and F. A. Schaposnik, Phys. Rev. Lett.56, 2564 (1986); Phys. Rev. D 34, 3206(1986) .
\bibitem{pd43}N.Manton, Nucl. Phys. B 400, 624(1993).
\bibitem{duth}Y. Q. Wang, T. Y. Si, Y. X. Liu and Y.S. Duan, Mod. Phys. Lett. A 20 3045(2005)
              [hep-th/0508111]
\bibitem{pa01}P. A. Horv$\acute{\mathrm{a}}$thy, J.-C. Yera, Lett. Math. Phys.
              46, 111-120 (1998) [hep-th/9805161].
\bibitem{pa02}P. A. Horv$\acute{\mathrm{a}}$thy,  Lett. Math. Phys. 49, 67-70 (1999) [hep-th/9903116].
\bibitem{dg01}Y. S. Duan, M. L. Ge, Sci. Sin. 11, 1072(1979); Y. S. Duan and X. H. Meng, J. Math.Phys. 34, 1149 (1993).
\bibitem{du01}Y. S. Duan, G. H. Yang, and Y. Jiang, Gen. Rel. Grav. 29, 715 (1997)
\bibitem{sa01}Y. S. Duan: SALC-PUB-3301(1984).
\bibitem{li04}Y. S. Duan and X. G. Lee, Helv.Phys.Acta. 58, 513 (1995).
\bibitem{li05}X. G. Lee, M. Baldo and Y. S. Duan, Gen. Rel. Grav. 29,
              715 (1997).
\bibitem{we01}X. G. Lee, et. al. Commom. Theor. phys.(accepted).arxiv:hep-th/0604138
\bibitem{pd04}H. Hopf, Math. Ann. 96, 209 (1929).
\bibitem{delt}A. S. Achwarz, Topology for Physicists. Springer-Verlag,
Berlin,1994.

\bibitem{book}B. A. Dubrovin, A. T. Fomenko, S. P. Novikov, \emph{The
              Geometry and Toplogy of Manifolds}.(Springer-Verlag,
              Berlin, 1985)
\end{thebibliography}
\end{document}